# An electron-spin qubit platform assembled atom-by-atom on a surface

Yu Wang[1,2,†], Yi Chen[1,2,†], Hong T. Bui[1,3,†], Christoph Wolf[1,2], Masahiro Haze[1,4], Cristina Mier[1,5], Jinkyung Kim[1,3], Deung-jang Choi[1,5,6,7], Christopher P. Lutz[8], Yujeong Bae[1,3,*], Soo-hyon Phark[1,2,*], Andreas J. Heinrich[1,3,*]

[1]Center for Quantum Nanoscience, Institute for Basic Science (IBS), Seoul 03760, Korea

[2]Ewha Womans University, Seoul 03760, Korea

[3]Department of Physics, Ewha Womans University, Seoul 03760, Korea

[4]The Institute for Solid State Physics, University of Tokyo, Kashiwa 277-8581, Japan

[5]Centro de Física de Materiales CFM/MPC (CSIC-UPV/EHU), 20018 Donostia-San Sebastián, Spain

[6]Donostia International Physics Center (DIPC), 20018 Donostia-San Sebastián, Spain

[7]Ikerbasque, Basque Foundation for Science, 48013 Bilbao, Spain

[8]IBM Research Division, Almaden Research Center, San Jose, CA 95120, USA

*Corresponding authors. Email: bae.yujeong@qns.science, phark@qns.science, heinrich.andreas@qns.science

†These authors contributed equally to this work.

**Creating a quantum-coherent architecture at the atomic scale has long been an ambition in quantum science and nanotechnology[1,2]. This ultimate length scale requires the use of fundamental quantum properties of atoms, such as the spin of electrons, which naturally occurs in many solid-state environments and allows high-fidelity operations and readout by electromagnetic means[3-5]. Despite decades of effort, however, it remains a formidable task to realize an atomic-scale quantum architecture where multiple electron spin qubits can be**

**precisely assembled, controllably coupled, and coherently operated. Electron spin qubits created in dopants in semiconductors and color centers in insulators, for example, can be well controlled individually[6-8] but are difficult to couple together into a circuit. On the other hand, multiple magnetic atoms and molecules on surfaces can be coupled to each other by building sophisticated atomic structures using a scanning tunneling microscope (STM)[9-13], but coherent operation has so far been limited to a single qubit in the tunnel junction[14,15]. Here we demonstrate an atomic-scale qubit platform by showing atom-by-atom construction, coherent operations, and readout of multiple electron-spin qubits on a surface. To enable the coherent control of "remote" qubits that are outside the tunnel junction, we complement each electron spin with a local magnetic field gradient from a nearby single-atom magnet[16,17]. To enable readout of remote qubits, we employ a sensor qubit in the tunnel junction and implement pulsed double electron spin resonance. Using these methods, we demonstrate fast single-, two-, and three-qubit operations in an all-electrical fashion. Our work marks the creation of an Angstrom-scale qubit platform, where quantum functionalities using electron spin arrays, built atom-by-atom on a surface, are now within reach.**

Constructing and coherently controlling novel nanoscale qubit systems lie at the heart of quantum-coherent nanoscience[18-20]. An attractive approach is to employ atomic-level fabrication by a scanning tunneling microscope (STM), whereby atom manipulation[9,21,22] or selective desorption[6,23] can lead to designed quantum spin architectures. As a first step towards *in situ* operation of atomic quantum devices, a radio-frequency (RF) voltage has been used to coherently control a single electron spin in the STM tunnel junction[14,15], whose spin states can be read out through spin-polarized tunnel current[10,12,24-27] in a so-called ESR-STM setup (ESR: electron spin resonance)[28]. However, harnessing the quantum functionalities of this platform requires multiple addressable qubits that lie outside the sub-nanometer tunnel junction region.

Figure 1a illustrates our strategy to construct such an atomic-scale multi-qubit platform. Addressable "remote" qubits are created on a bilayer MgO film by positioning a spin-1/2 hydrogenated Ti atom[11,29] ~0.6 nm away from a single-atom magnet (Fe)[30] using STM-based atom manipulation (see Methods). Fe atoms are used here to supply a local magnetic field gradient, which can convert a tip-induced radio-frequency (RF) electric field into an effective driving magnetic field[31,32], in analogy to micromagnet driving in quantum dots[33]. These qubits remain remote from the STM tunnel junction to avoid decoherence induced by the tunnel current. To enable readout, a sensor qubit consisting of a Ti atom is positioned in the tunnel junction and is weakly coupled to the remote qubits. Initialization is performed thermally by cooling the sample to 0.4 K and applying an external magnetic field, producing a thermal spin state having a predominant population in the spin ground state (~90% under a 0.7 T field).

A representative structure composed of two remote qubits and one sensor qubit is constructed atom-by-atom using atom manipulation (see Methods), as shown in Fig. 1b,c. The qubit-qubit couplings are sensitive to their atomic separations down to the Angstrom level[11,12,29] (Extended Data Figs. 1 and 2), consistent with results in other material systems[34]. This atomically-precise construction scheme thus allows us to engineer the resonant frequencies and couplings among all the spins, an essential step for addressing and detecting the multiple qubits individually (see Methods).

Detection of the remote qubits is achieved through ESR spectroscopy of the sensor qubit, whose ESR transition frequency depends on the quantum states of other qubits (Fig. 1d,e). The ESR frequencies of the qubits are designed to be sufficiently separated (see Methods and Extended

Data Fig. 1) that the multi-qubit states are well described by Zeeman product states, where we use blue (red) brackets to denote the quantum states of the sensor (remote) qubit. In a structure composed of a sensor qubit and a remote qubit (Fig. 1d), the two ESR transitions of the sensor qubit at frequencies $f_1$ (corresponding to transition $|0\rangle|0\rangle \leftrightarrow |1\rangle|0\rangle$) and $f_2$ ($|0\rangle|1\rangle \leftrightarrow |1\rangle|1\rangle$) detect the populations of $|0\rangle$ and $|1\rangle$ states of the remote qubit, respectively. Multiple remote qubits can be simultaneously sensed in a similar fashion (Fig. 1e).

To demonstrate this qubit control and readout scheme, we begin by measuring a remote qubit's ESR spectrum with the tip positioned above the sensor qubit, as sketched in Fig. 2a. The energy levels and ESR transitions of this two-qubit structure are illustrated in Fig. 2b and its STM image is shown in Fig. 1b. In order to individually address the sensor and remote qubits, we use two RF sources to apply two consecutive RF voltage pulses to the STM tip. A control pulse is applied at frequency $f_R$ to control the remote qubit, followed by a sensing pulse at frequency $f_S$ acting on the sensor qubit (Fig. 2b,c). A sensing pulse applied at $f_S = f_1$ ($f_2$) results in an ESR signal that depends on the population of state $|0\rangle$ ($|1\rangle$) of the remote qubit. To obtain the ESR spectrum of the remote qubit, we sweep the frequency $f_R$ of the control pulse across the remote qubit resonances while keeping the sensing pulse fixed at a resonance frequency of the sensor qubit. When $f_R$ matches an ESR transition of the remote qubit (e.g., $|0\rangle|0\rangle \leftrightarrow |0\rangle|1\rangle$), the joint two-qubit state populations are altered, resulting in a detectable decrease (increase) of the sensor's ESR signal at $f_1$ ($f_2$) (Fig. 2d and Methods). This measurement is conceptually similar to ensemble double electron spin resonance spectroscopy[35] and allows us to directly obtain the resonance frequencies of the remote qubit ($f_3$ and $f_4$, see Fig. 2d) even though no tunnel current passes through it.

Single-qubit control is performed in our platform by coherently driving a qubit independent of other qubits' quantum states. This is achieved by exciting all ESR transitions of a remote qubit with the same driving strength[14], as illustrated in the insets of Fig. 2e,f. The coherent rotation of a remote qubit on its Bloch sphere can be conducted by varying the pulse duration $\tau_R$ of the control pulse. A subsequent sensing pulse at $f_1$ reveals coherent Rabi oscillations of the remote qubit (Fig. 2e).

Two-axis control[36] of the remote qubit is demonstrated by varying the relative phase φ of two π/2-pulses (Fig. 2f). On the Bloch sphere in the rotating frame, the first π/2-pulse rotates the qubit from the *z*-axis to the *y*-axis, and the second performs a π/2-rotation around an axis in the *x*-*y* plane at an angle of φ to the *x*-axis (lower inset, Fig. 2f). The resulting signal is well described by the expected cos(φ) dependence (Fig. 2f)[36]. Two-axis control allows arbitrary single-qubit operations as discussed in Ref. [37].

We next discuss two-qubit operations. Controlled NOT (CNOT) operations are straightforward to carry out in our platform by selectively exciting ESR transitions that correspond to a specific quantum state of a control qubit. In the case of a two-qubit structure, a CNOT operation of the remote qubit can be performed by driving a single ESR transition such as $|0\rangle|0\rangle \leftrightarrow |0\rangle|1\rangle$ at $f_R = f_3$ (Fig. 3a), which inverts the remote qubit only if the sensor qubit is in state $|0\rangle$. The quantum state of the remote qubit is then measured by applying a sensing pulse at $f_1$ (Fig. 3a). Sensing at $f_2$ (instead of at $f_1$) shows oscillations of the opposite sign because this effectively detects the population of state $|1\rangle$ (instead of state $|0\rangle$) of the remote qubit (Extended Data Fig. 4)[38]. The oscillations in Fig. 3a show a CNOT operation time of ~13 ns for the remote qubit. The operation rate increases in proportion to the RF amplitude (Fig. 3b upper, Extended Data Fig. 3a). This fast operation results from the strong magnetic field gradient of the neighboring Fe atom. In contrast,

ESR of the sensor qubit, as well as previous ESR-STM spectroscopy[11,27,32], relies on the tip's magnetic field gradient. A consequence of driving by the surface Fe is that the CNOT rate is independent of time-average tunnel current $I_{DC}$, which corresponds to the tip's proximity (Fig. 3b lower, Extended Data Fig. 3b).

To evaluate the effect of the CNOT operation on the remote qubit, we perform controlled rotations of the sensor qubit without and with this CNOT operation (Fig. 3c,d). Since the initial thermal population of the two-qubit system is predominantly in state $|0\rangle|0\rangle$, a controlled rotation of the sensor qubit at $f_1$ starting from the initial state $|0\rangle|0\rangle$ (i.e., without a CNOT operation) shows clear oscillations because a pulse at $f_1$ excites the transition $|0\rangle|0\rangle \leftrightarrow |1\rangle|0\rangle$. Whereas measurement at transition $f_2$ shows no obvious oscillations[14] (Fig. 3c) because remote qubit state $|1\rangle$ is nearly unoccupied. These coherent oscillations are reversed after a CNOT operation of the remote qubit at $f_3$ (Fig. 3d), which transfers the predominant population from state $|0\rangle|0\rangle$ to state $|0\rangle|1\rangle$, hence becoming accessible to a sensing pulse at $f_S = f_2$ (i.e., $|0\rangle|1\rangle \leftrightarrow |1\rangle|1\rangle$) but not $f_1$[39]. We further verify that a CNOT operation of the remote qubit at $f_4$ (instead of $f_3$) does not significantly affect the oscillations observed on the sensor qubit (Extended Data Fig. 6), highlighting the selective, controlled nature of CNOT operations.

Although single- and two-qubit operations are sufficient to generate arbitrary quantum circuits[40], multi-qubit operations can significantly reduce the execution time and mitigate accumulated operation errors[41,42]. As long as the qubit-qubit couplings can be spectroscopically resolved, our platform allows fast, native qubit operations having multiple control qubits (by selectively exciting certain ESR transitions). To demonstrate this ability, we construct a three-qubit structure composed of two remote qubits (referred to as RQ1 and RQ2) and a sensor qubit (as sketched in Fig. 4a and imaged in Fig. 1c). We selectively drive RQ1's $|0\rangle|00\rangle \leftrightarrow |0\rangle|10\rangle$ transition (Extended Data Fig. 7a), corresponding to a rotation of RQ1 if and only if the sensor qubit and RQ2 are both in state $|0\rangle$ (Fig. 4a). When varying the pulse duration $\tau_R$, this controlled-controlled operation is seen to cause oscillating populations between states $|00\rangle$ and $|10\rangle$ of the two remote qubits, whereas populations of states $|01\rangle$ and $|11\rangle$ are left unchanged (Fig. 4b). Here the changes of populations are determined from the intensities of the four ESR transitions of the sensor qubit (blue pulse in Fig. 4a). Similarly, the controlled-controlled operation $|0\rangle|01\rangle \leftrightarrow |0\rangle|11\rangle$ is seen to cause oscillating populations between states $|01\rangle$ and $|11\rangle$ of the two remote qubits, albeit with reduced amplitudes due to reduced initial thermal populations in these states (Extended Data Fig. 7e). From these measurements we obtain a controlled-controlled NOT (CCNOT) operation time as short as 20 ns. Single-, two-, and multi-qubit gates in our platform have comparably fast operation times because they only differ in the frequency content of the RF pulses[43,44] (Extended Data Fig. 8).

In STM-based approaches, tunneling electrons have posed severe limitations on the energy relaxation time $T_1$ and coherence time $T_2$ of the spins[30,45]. This limitation is overcome in our scheme because tunneling electrons pass only through the sensor qubit and do not flow through the remote qubits (Fig. 1a). To characterize the remote qubits, we first perform an inversion recovery measurement[46] to obtain an energy relaxation time $T_1$ = 166 ± 14 ns[12]. A pronounced improvement is seen in the quantum coherence of remote qubits, which is already visible in the higher quality of the remote qubit's Rabi oscillations (Fig. 3a) compared to the sensor qubit (Fig. 3c and 3d). Figure 5b shows the Ramsey signal measured at a tunnel current of $I_{DC}$ = 10 pA, from which we extract a coherence time $T_2^*$ = 86 ± 13 ns. We find that $T_2^*$, the coherence time subject to inhomogeneous broadening, depends on the tunnel current and hence on the tip height (Fig. 5b inset, Extended Data Fig. 5a), suggesting that the proximity of the STM tip to the remote qubits influences their quantum coherence time despite the absence of tunnel current. This decoherence effect is likely due to field fluctuations arising from slight, uncontrolled tip motions[45]. To cancel the effect of this inhomogeneous broadening, we perform a spin-echo measurement and observe that the measured $T_2^{\mathrm{Echo}}$ shows negligible dependence on the tunnel current (Fig. 5c inset and Extended Data Fig. 5b). The measured coherence time $T_2^{\mathrm{Echo}} = 300 \pm 54$ ns (Fig. 5c) strikingly approaches the theoretical limit of $2T_1$. These measurements highlight that the quantum coherence of remote qubits after spin-echo filtering is limited by energy relaxation events, in contrast to other solid-state qubits, where $T_1 \gg T_2^{\mathrm{Echo}}$ usually applies.

Further increases in $T_1$ and $T_2^{\mathrm{Echo}}$ should be possible by use of thicker insulating layers to better isolate the qubits from substrate electrons and by switching to longer-lived single-atom magnets such as Ho and Dy[16,17,47]. The remote qubits can in principle be positioned on a fully insulating material as long as they couple to the sensor spin placed on a weakly conductive region of the surface. A variety of material systems, such as those involving two-dimensional heterostructures, offer such possibilities[48,49].

We have shown the atom-by-atom fabrication and coherent manipulation of a novel electron-spin qubit platform. Coherently controlling multiple qubits with a single STM tip overcomes a previous limitation in STM-based quantum control schemes. A unique benefit of our surface-based electron-spin approach is the myriad of available spin species[28,50] and the vast variety of two-dimensional geometries that can be controllably assembled. We anticipate that multi-spin entanglement demonstration[51], quantum computation[52], sensing[53] and simulation[54] protocols using an atomic architecture on a surface can now be attempted.

**Acknowledgments**

We thank A. Ardavan, M. Ternes, J. F. Rossier, D. Loss, F. Donati, and F. Cho for fruitful discussions. This work was supported by the Institute for Basic Science (IBS-R027-D1). C.P.L. acknowledges support from the Office of Naval Research, grant N00014-21-1-2467. M.H. acknowledges financial support from University of Tokyo Global Activity Support Program for Young Researchers (FY2020). D.-J.C. and C.M. acknowledge funding from the Spanish State Research Agency grants (RTI2018-097895-B-C44, Excelencia EUR2020-112116 and PID2021-127917NB-I00) funded by MCIN/AEI/10.13039/501100011033 and by "ERDF A way of making Europe", and Eusko Jaurlaritza (project PIBA 2020_1_0017).

**Author contributions**

A.J.H., S.P., and C.P.L. conceived the project. Y.W., Y.C., H.T.B., M.H., C.M., J.K., D.-J.C., Y.B., and S.P. performed the measurements and data analysis. C.W. carried out simulations to optimize the measurement schemes. All authors discussed and prepared the manuscript together.

**Competing interests**

The authors declare no competing interests.

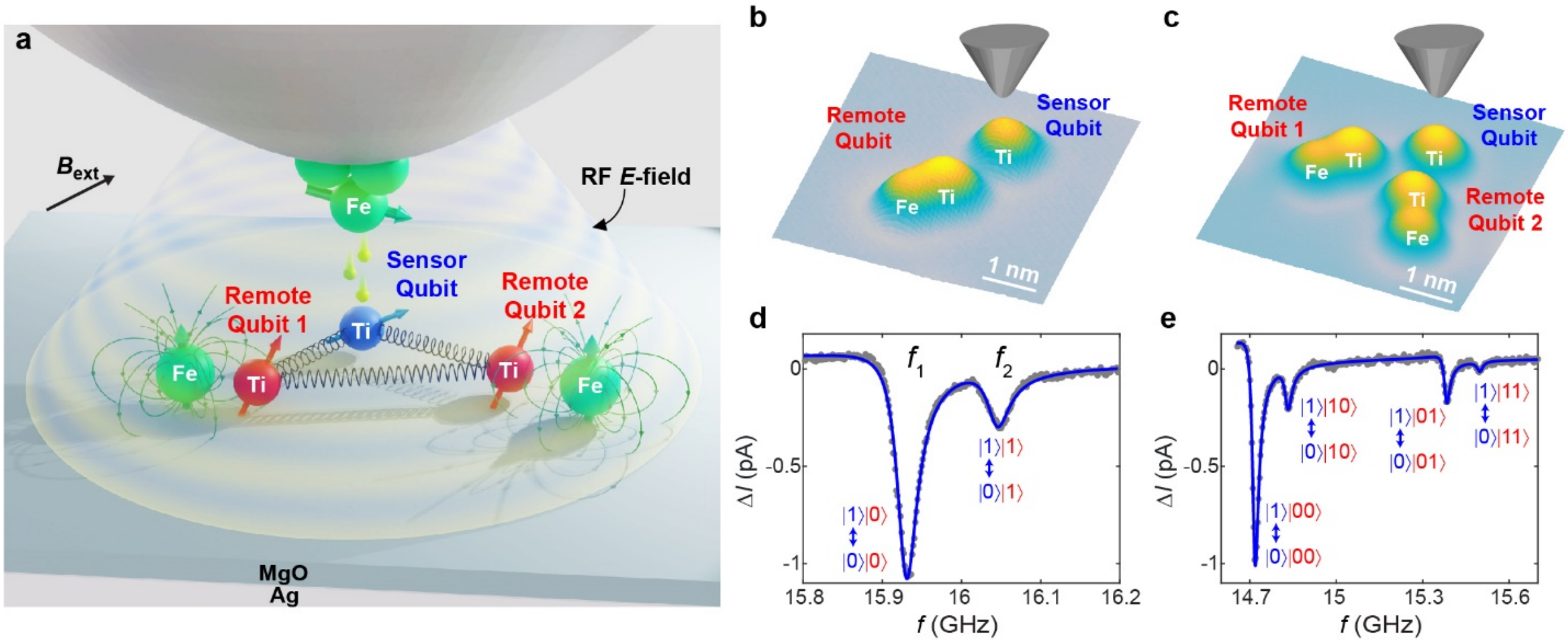

**Fig. 1 | Bottom-up construction of multiple coupled electron spin qubits. a,** Schematic: A sensor spin qubit (Ti, blue) is placed under the apex of a spin-polarized STM tip for readout. Remote qubits are constructed at precise separations to the sensor qubit by atom manipulations. Each remote qubit is composed of a spin-1/2 Ti atom (red) and a single-atom magnet (Fe) (green), where Fe's magnetic field gradient, in combination with the RF electric field between the tip and the sample, coherently drives remote qubits. **b**,**c**, Constant-current STM images showing atom-by-atom construction of a multi-qubit structure (image size: 5.0 nm × 5.0 nm). **d,e**, Continuous-wave ESR spectra measured with the tip positioned on the sensor qubit in the structures shown in **b** and **c**. Each ESR resonance of the sensor qubit distinguishes a quantum state of the remote qubits (red kets). The quantum states of the sensor qubit are labeled by blue kets. Imaging conditions in **b** and **c**: sample bias voltage $V_{DC}$ = 100 mV, time-averaged tunnel current $I_{DC}$ = 10 pA. ESR conditions in **d** and **e**: $V_{DC}$ = 50 mV, $I_{DC}$ = 20 pA, zero-to-peak RF voltage $V_{RF}$ = 30 mV.

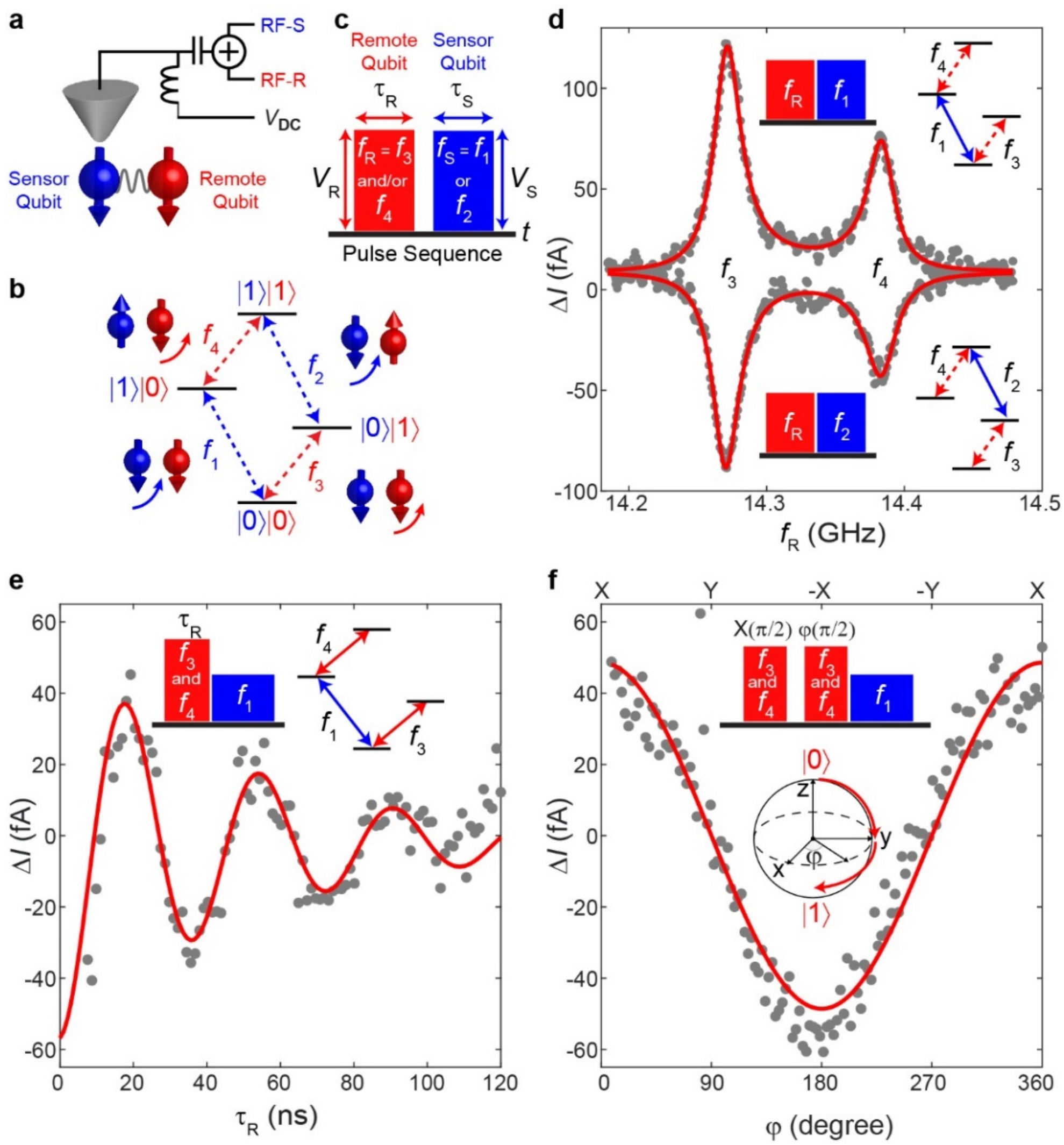

**Fig. 2 | Coherent control of a single remote qubit. a,** Schematic of the measurement scheme. Two RF tones, RF-S and RF-R, are applied to the STM tip for the coherent control of the sensor and remote qubit, respectively. The atomic structure used in this figure is shown in Fig. 1b. **b,** Energy diagram and ESR transitions of the two-qubit system. **c,** Typical pulse sequence composed of a control pulse on the remote qubit (red) followed by a sensing pulse on the sensor qubit (blue). The control pulse can induce conditional or unconditional qubit rotations depending on its frequency content. **d,** ESR spectra of the remote qubit measured with the tip positioned on the sensor qubit. The two curves in **d** correspond to different sensing frequencies (upper: $f_S = f_1$, lower: $f_S = f_2$). Insets: pulse sequence and ESR transitions involved in each spectrum. **e,** Rabi oscillations of the remote qubit performed by simultaneously driving $f_R = f_3$ and $f_4$ to induce unconditional qubit rotations. The sensing pulse is subsequently applied at $f_S = f_1$. Red curve is a fit to an exponentially decaying sinusoid. **f,** Two-axis control of the remote qubit shown by sweeping the relative phase φ between two π/2 pulses ($f_R = f_3$ and $f_4$; $f_S = f_1$). Solid curve shows a cosine fit. The trajectory of the remote qubit is illustrated by red arrows on the Bloch sphere in the rotating frame. ESR conditions in **d**: $V_{DC}$ = 50 mV, $I_{DC}$ = 20 pA, $V_S = V_R$ = 30 mV, $\tau_R = \tau_S$ = 200 ns. **e** and **f**: $V_{DC}$ = 50 mV, $I_{DC}$ = 20 pA, $V_S$ = 50 mV, $V_R$ = 120 mV, $\tau_S$ = 200 ns.

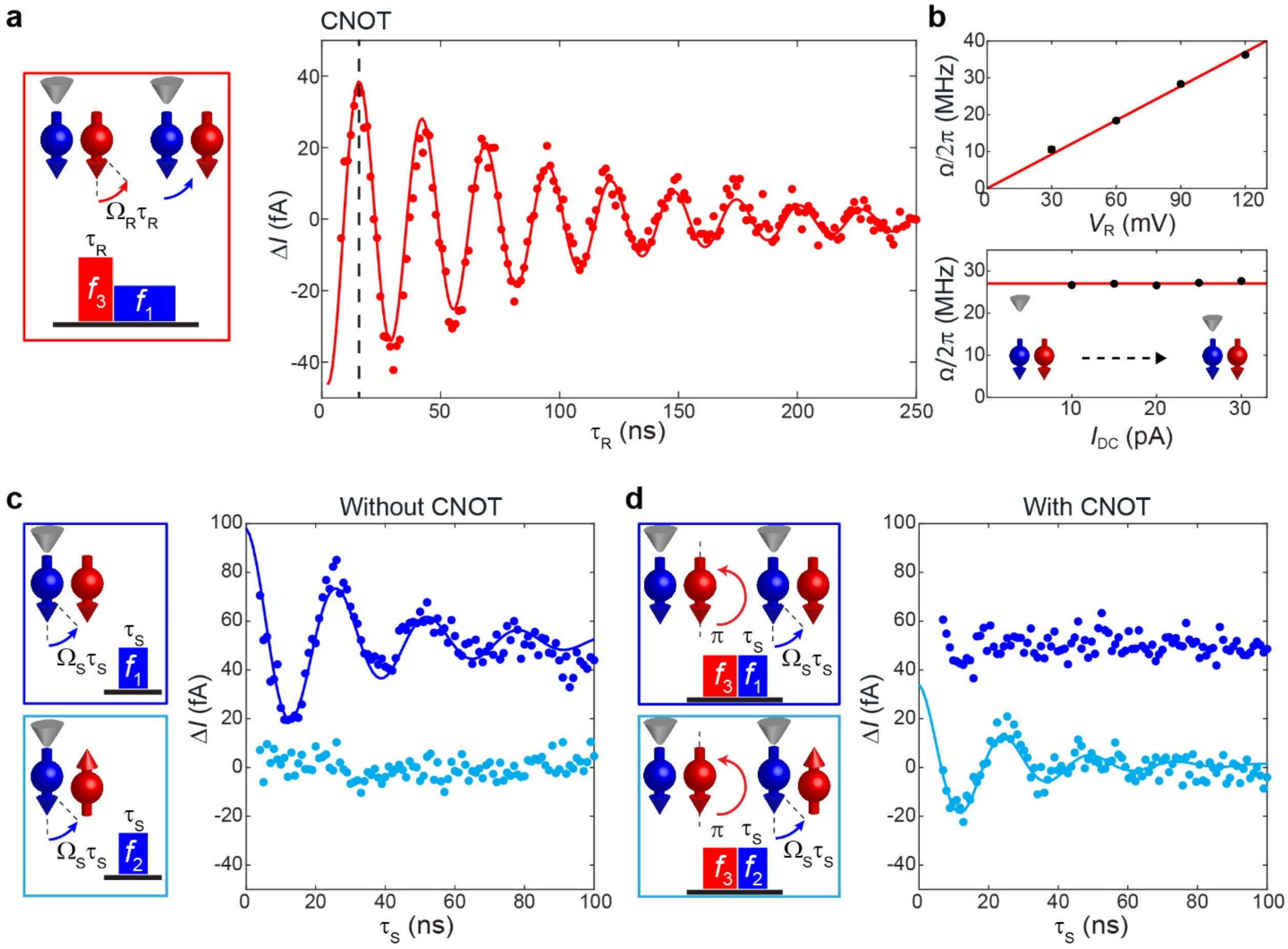

**Fig. 3 | Two-qubit operations in the coupled-qubit structure shown in Fig. 1b. a,** Controlled rotation of the remote qubit, conditional on the sensor qubit state being $|0\rangle$, obtained by driving the transition $|0\rangle|0\rangle \leftrightarrow |0\rangle|1\rangle$ at $f_R = f_3$. Red points show coherent oscillations of the remote qubit detected through a sensing pulse at $f_S = f_1$, as shown schematically on the left. Red curve is a decaying sinusoidal fit. The CNOT operation time is ~13 ns (dotted line). **b,** The rate of controlled rotations, $\Omega/2\pi$, of the remote qubit. The rate increases linearly with the RF voltage $V_R$ (upper) but remains unchanged as the tunnel current $I_{DC}$ (lower) is varied. Error bars are comparable with the size of symbols. Solid curves are linear fits. Inset illustrates the change of tip-sensor separations. **c**,**d**, Controlled rotations of the sensor qubit without (c) and with (d) a CNOT operation at $f_R = f_3$ on the remote qubit. Blue (cyan) points measured with $f_S = f_1$ ($f_2$) show sensor Rabi oscillation (solid lines are decaying exponential fits) only when the remote qubit has a significant population in state $|0\rangle$ ($|1\rangle$). The CNOT operation transfers the predominant population from state $|0\rangle|0\rangle$ to state $|0\rangle|1\rangle$, thus causing the opposite trends of the oscillations in **c** and **d**. Blue curves are shifted vertically by 50 fA for clarity. ESR conditions in **a**: $V_{DC}$ = 50 mV, $I_{DC}$ = 10 pA, $V_S$ = 30mV, $V_R$ = 120 mV, $\tau_S$ = 200 ns. **c**: $V_{DC}$ = 20 mV, $I_{DC}$ = 7.5 pA, $V_S$ = 100 mV. **d**: $V_{DC}$ = 20 mV, $I_{DC}$ = 7.5 pA, $V_S = V_R$ = 100 mV.

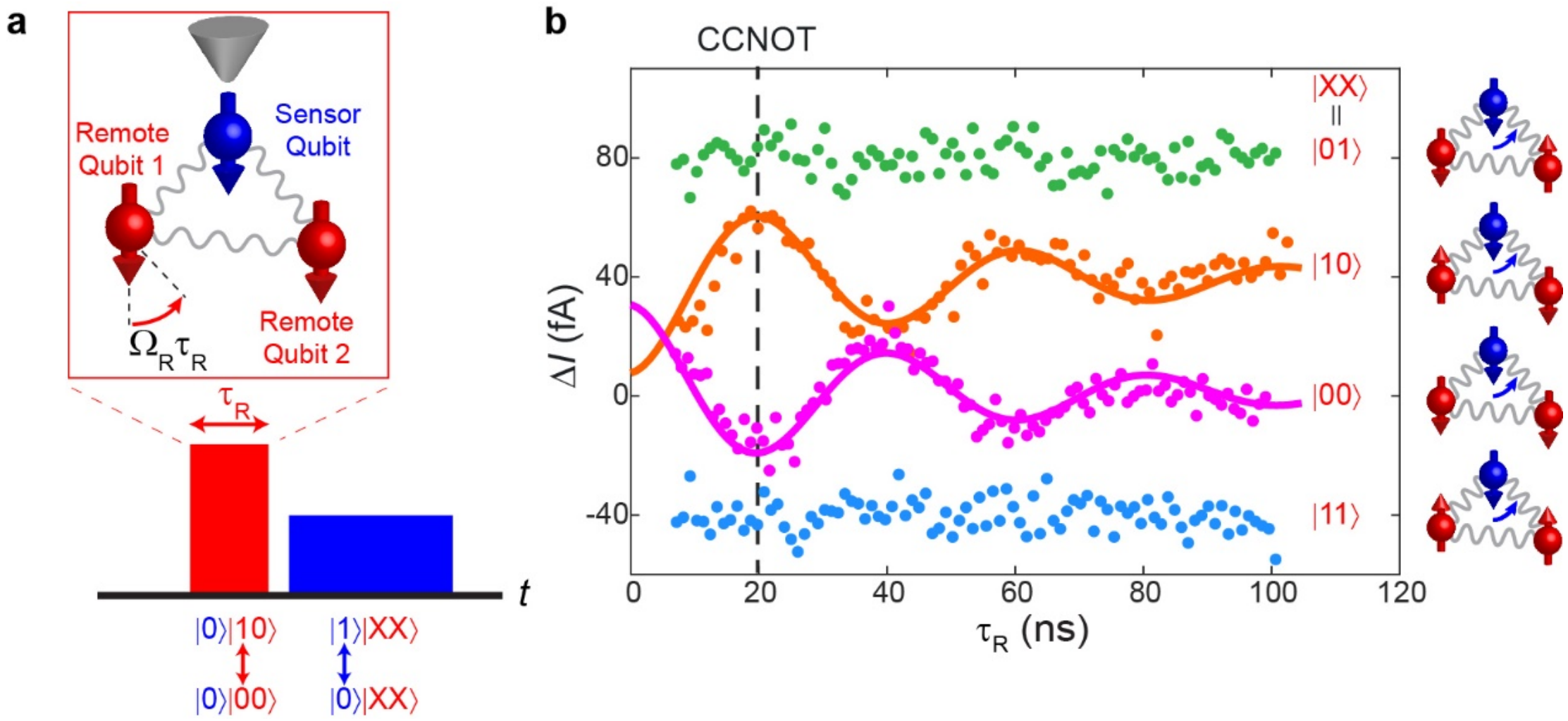

**Fig. 4 | Three-qubit operations in a multi-qubit atomic structure. a,** Schematic showing the control scheme of a multi-qubit structure (Fig. 1c) composed of two remote qubits and a sensor qubit. **b,** Controlled-controlled operation of remote qubit 1 performed by driving the transition |0⟩|00⟩ ↔ |0⟩|10⟩ (red pulse in **a**). Four different frequencies of the sensor qubit are used for sensing the four remote qubit states (blue pulse in **a**, where the detected states of the remote qubits are labelled in red kets in **b**). The two oscillating curves in **b** (orange and magenta) correspond to detection of the remote qubits' |00⟩ or |10⟩ states upon the controlled-controlled operation at |0⟩|00⟩ ↔ |0⟩|10⟩. The CCNOT operation time is ~20 ns (dotted line). The other two qubit states are nearly unoccupied so they show no clear oscillations. Green, orange and blue curves are vertically shifted for clarity. ESR conditions in **b**: $V_{DC}$ = 50 mV, $I_{DC}$ = 20 pA, $V_S$ = 50 mV, $V_R$ = 80 mV, $\tau_S$ = 200 ns.

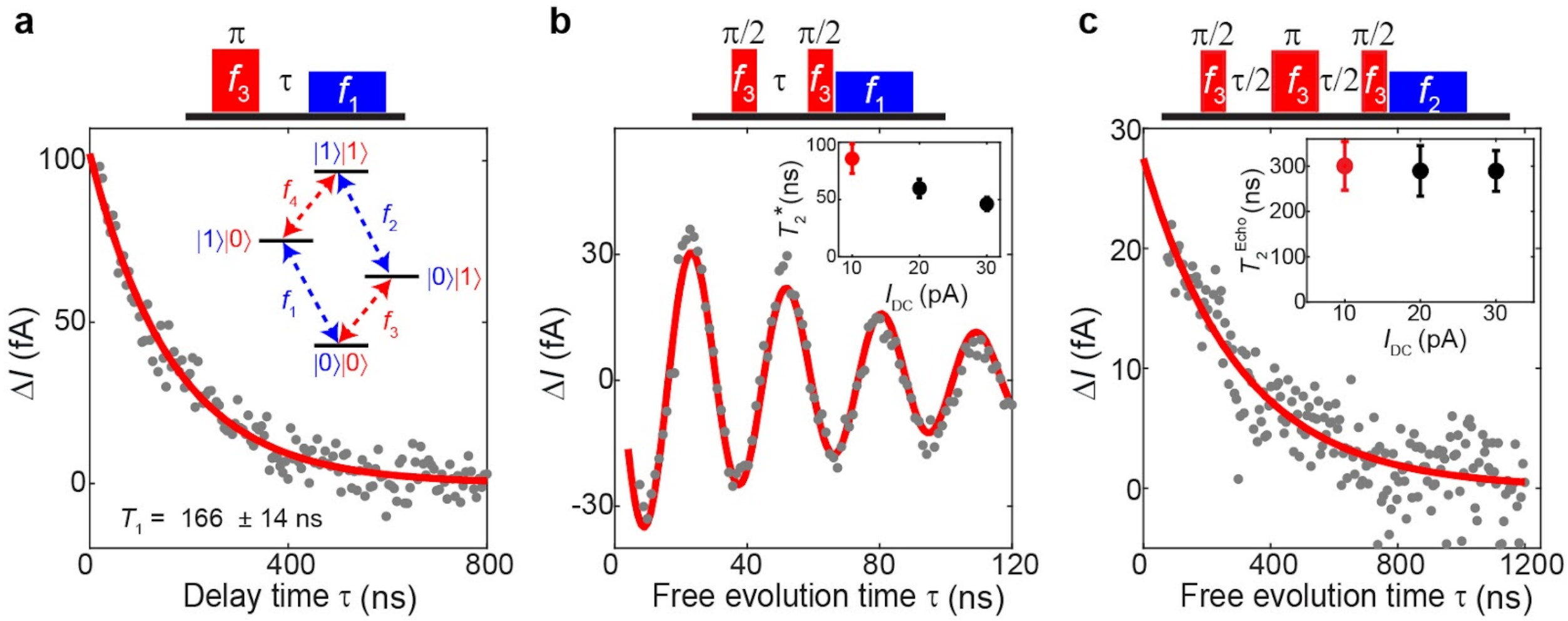

**Fig. 5 | Relaxation and coherence properties of remote qubits.** The two-qubit structure shown in Fig. 1b is used in these measurements. **a,** Relaxation time $T_1$ of the remote qubit measured by an inversion recovery scheme. The exponential fit (solid line) yields a relaxation time $T_1$ of 166 ± 14 ns for the remote qubit. Inset: Labels of ESR transitions used in this figure (same as Fig. 2b). **b,** Ramsey measurements of the remote qubit with $f_S = f_1$ and $f_R = f_3 + 30$ MHz yielding a coherence time $T_2^* = 86 \pm 13$ ns. Inset: $T_2^*$ shows a weak dependence on the tunnel current $I_{DC}$. **c,** Spin-echo measurements of the remote qubit measured with $f_S = f_2$ and $f_R = f_3$. The exponential fit (solid line) yields a coherence time of $T_2^{Echo} = 300 \pm 54$ ns. Inset: $T_2^{Echo}$ shows no dependence on $I_{DC}$. ESR conditions in **a**: $V_{DC} = 50$ mV, $I_{DC} = 20$ pA, $V_S = 50$ mV, $V_R = 120$ mV, $\tau_S = 200$ ns. **b** and **c**: $V_{DC} = 50$ mV, $I_{DC} = 10$ pA, $V_S = 60$ mV, $V_R = 120$ mV, $\tau_S = 200$ ns.

## Methods

### Experimental setup

Experiments were performed with a commercial STM (Unisoku, USM1300) operating at 0.4 K. Two radio frequency (RF) generators (Agilent E8257D and E8267D) were combined by an RF power combiner. The combined RF signal was added to the DC bias voltage $V_{DC}$ through a diplexer at room temperature and then applied to the STM tip. Both RF generators were gated by an arbitrary waveform generator (AWG, Tektronix 5000). The gating of the vector RF generator (E8267D) was performed by connecting two analog output channels of the AWG to the generator's in-phase and quadrature (I and Q) ports in order to control the phase of the RF pulses and drive multiple frequencies. The tunnel current was sensed by a room-temperature electrometer (Femto DLPCA-200). During ESR measurements, the signals were chopped at 95 Hz and sent to a lock-in amplifier (Stanford Research Systems SR860) and recorded (National Instruments 6363). A vector magnetic field (0.6 – 0.7 T) was applied mainly in the sample plane, with an out-of-plane component adjusted between 0.1 and 0.3 T. The bias voltage $V_{DC}$ refers to the sample voltage relative to the tip. The STM constant-current feedback loop was set to a low gain during the measurements.

### Measurement schemes and data processing

Each data point in the figures corresponds to a data acquisition time of ~1–2 s, during which the pulse sequences were unchanged and the averaged output voltage from the lock-in amplifier was recorded. Each lock-in cycle (~10.5 ms) was divided into an A-subcycle (~5.25 ms) and a B-subcycle (~5.25 ms). Each pulse sequence was set to be roughly 1 μs long to allow enough time for qubit relaxation before the start of the next sequence[14]. As a result, each lock-in subcycle was filled with repeated identical sequences with roughly 5000 repetitions. The pulse sequences used in A-subcycles are illustrated in the figures. Schemes in B-subcycles are chosen to cancel linear backgrounds in the raw data, which are caused by RF-voltage-induced broadening of a nonlinear d*I*/d*V* spectrum (RF rectification effects)[55]. Continuous-wave ESR spectra (Figs. 1d, 1e and Extended Data Figs. 1 and 2) as well as sensor qubit rotations (Fig. 3c) were collected with the RF voltage applied only in the A-subcycles and absent in the B-subcycles.

In this work, the remote measurement scheme reads out the *z*-projection of the quantum state of the remote qubits by detecting conditional spin rotations of the sensor qubit.

Coherent oscillations and Ramsey signals were fitted with a decaying sinusoidal function after subtracting a linear background. A constant current offset induced by remaining RF rectification effects was removed in Figs. 2f, 5a and 5c.

**Sample preparation**

An atomically clean Ag(100) substrate was prepared by alternating Ar+ sputtering and annealing cycles. MgO films were grown on the Ag substrate at 700 K by evaporating Mg in an $O_2$ atmosphere of $1.1\times10^{-6}$ Torr. Fe and Ti atoms were deposited on pre-cooled MgO samples. All Ti atoms are believed to be hydrogenated[11,29]. All experiments were performed on bilayer MgO on Ag as confirmed by *I*-*z* spectroscopy on Fe atoms[30].

**STM tips used for data acquisition**

Magnetic tips used for measurements were made by picking up ~4–6 Fe atoms from the MgO surface until the tips yielded good ESR signals on Ti atoms. Data shown in this Article were taken with five different tips. These tips had different strength of spin polarization as shown in Extended

Data Fig. 5. Tip-1 was used to collect the data in Fig. 3b and Extended Data Fig. 3; tip-2 was used to collect the data in Figs. 1d, 2d, 3a, 5b, 5c and Extended Data Figs. 1, 4 and 9; tip-3 was used to collect the data in Figs. 2e, 2f and 5a; tip-4 was used to collect the data in Figs. 3c, 3d and Extended Data Fig. 6; tip-5 was used to collect the data in Figs. 1e, 4b and Extended Data Figs. 2 and 7.

**Atomic-scale fabrication of the qubit structures**

The atomic qubit structures were fabricated using a combination of atom manipulation techniques. Fe atoms were vertically manipulated by picking up and dropping off the atoms using a sharp STM tip. Ti atoms were horizontally manipulated by placing the tip in front of Ti and approaching the tip to the surface. All procedures are described in Refs. [28,56]. We found that the optimal distance between Ti and Fe atoms in a remote qubit is ~0.6 nm (Fig. 1b and remote qubit 1 in Fig. 1c). A closer distance causes modifications of local density of states of Ti atoms, while a longer distance greatly reduces the remote driving strength. Atomically identical copies of each structure were created and showed no significant differences in ESR measurements.

**Spin Hamiltonian**

The static Hamiltonian of the qubit system shown in Fig. 1d can be written as

$$H = g\mu_{\mathrm{B}}(\mathbf{B}_{\mathrm{Fe}} + \mathbf{B}_{\mathrm{ext}}) \cdot \mathbf{S}_{\mathrm{R}} + g\mu_{\mathrm{B}}\left(\mathbf{B}_{\mathrm{tip}} + \mathbf{B}_{\mathrm{ext}}\right) \cdot \mathbf{S}_{\mathrm{S}} + J_{\mathrm{S-R}}\, \mathbf{S}_{\mathrm{S}} \cdot \mathbf{S}_{\mathrm{R}}$$

$$+ D(\mathrm{S}_{\mathrm{S}z}\mathrm{S}_{\mathrm{R}z} - \mathbf{S}_{\mathrm{S}} \cdot \mathbf{S}_{\mathrm{R}})$$

where $\mathbf{S}_{\mathrm{S}}$ and $\mathbf{S}_{\mathrm{R}}$ are spin operators of the sensor and remote qubit, $J$ and $D$ are the exchange and dipolar coupling strengths, respectively. The couplings among sensor and remote qubits were engineered by tuning their atomic separations and orientations (Extended Data Fig. 2). To separate the qubits' resonance frequencies, three strategies were used:

1. Adjusting the effective field $\mathbf{B}_{\mathrm{Fe}}$ from Fe atoms on remote Ti atoms, which is determined by their atomic separations and orientations on the MgO lattice.

2. Using different bridge binding-site orientations of Ti atoms on the MgO lattice because Ti atoms in different binding-sites yield different *g*-factors[12,26].

3. Changing the amplitude and direction of the external magnetic field $\mathbf{B}_{\mathrm{ext}}$, which provides a global change of the qubit frequencies. In addition, different qubits' resonances shifted differently because of (1) their different *g* factors and (2) the presence of different $\mathbf{B}_{\mathrm{Fe}}$ (which is fixed in the out-of-plane direction[57] and barely affected by the external magnetic field).

**Data availability**

All data that support the plots within this paper and other findings of this study are available from the corresponding authors upon reasonable request.

**Code availability**

The codes used in this study are available from the corresponding authors upon reasonable request.

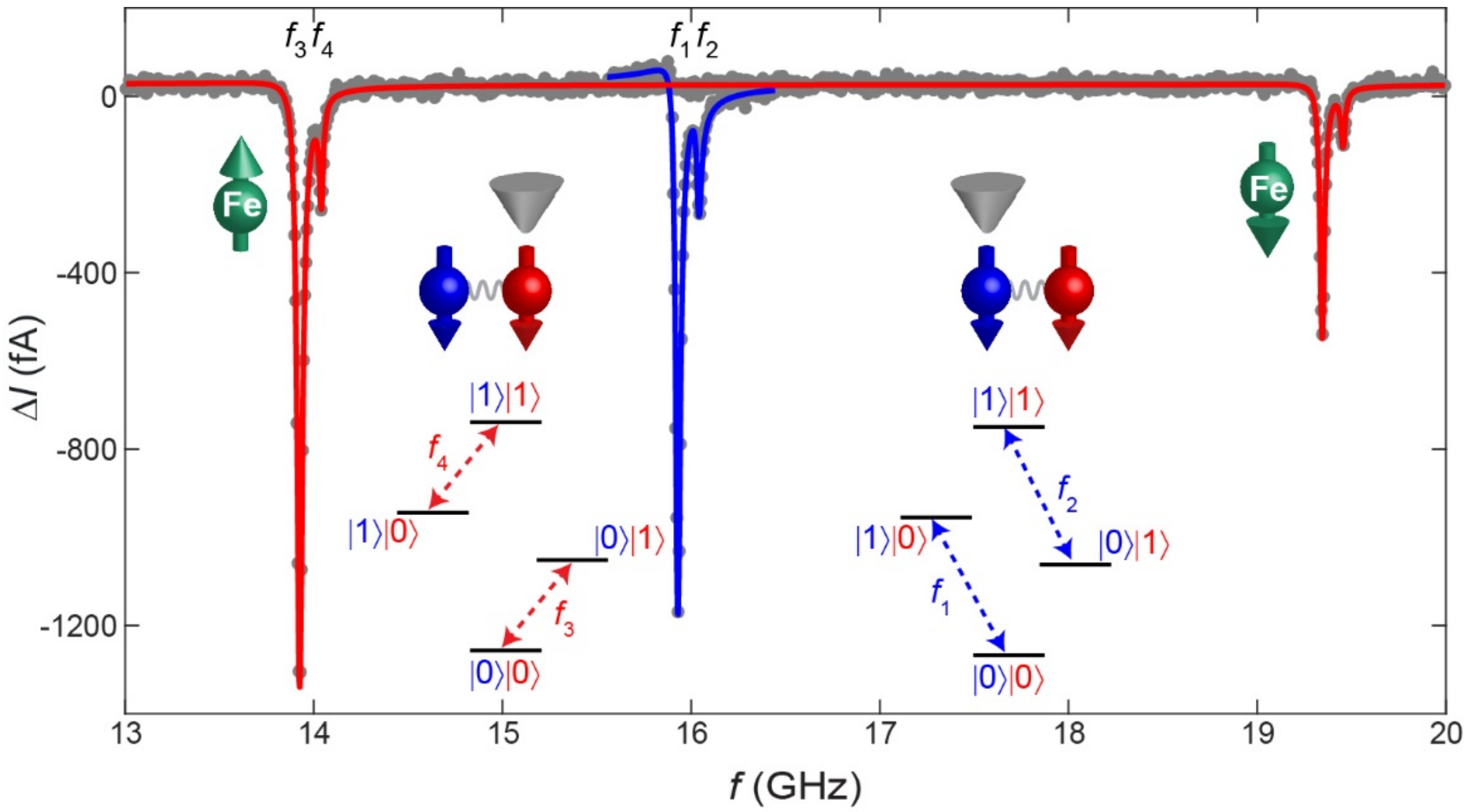

**Extended Data Fig. 1 | Wide-range continuous-wave ESR spectra measured in the two-qubit structure shown in Fig. 1b.** Unlike all spectra taken in the main text, where the STM tip is positioned only on the sensor qubit, the ESR spectra here are measured with the tip positioned either on the sensor qubit (blue) or on the "remote" qubit (red). The transitions corresponding to each ESR line are illustrated in the insets. Due to infrequent flipping of the Fe spin, the ESR spectrum measured on the remote qubit has two groups of peaks, corresponding to Fe's spin up or down states. In this work, only ESR peaks at $f_3$ and $f_4$ (corresponding to Fe in the ground state $|\uparrow\rangle$) are considered because of their predominant strength. Fe flipping is expected to affect the spectra only slightly because Fe's relaxation time is significantly longer than the timescale of each pulse sequence (1 μs)[1,2]. ESR conditions: $V_{DC}$ = 50 mV, $I_{DC}$ = 20 pA, $V_{RF}$ = 30 mV.

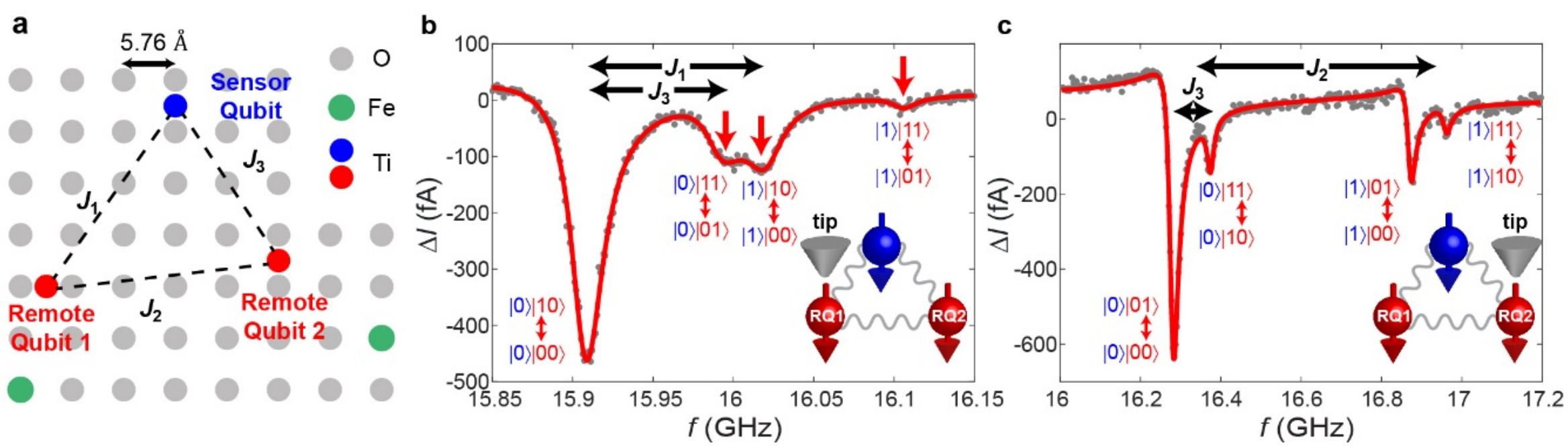

**Extended Data Fig. 2 | Atomic registry of Ti and Fe atoms on the MgO lattice and continuous-wave ESR spectra of the three-qubit structure shown in Fig. 1c. a,** Atomic registry of the three-qubit structure in Fig. 1c. The sensor Ti atom is shown in blue. In the remote qubits, Ti and Fe atoms are shown in red and green, respectively. Grey disks represent oxygen atoms. Fe atoms sit atop oxygen atoms, and Ti atoms sit at "bridge sites" between two oxygen atoms. The two-qubit structure in Fig. 1b has the same configuration but without remote qubit 2. **b,c,** Continuous-wave ESR spectra measured with the tip positioned on remote qubit 1 (RQ1) (b) and remote qubit 2 (RQ2) (c), as illustrated in the corresponding insets. The qubit-qubit interactions and transitions in each ESR spectrum are also labeled. From the ESR spectra we extract $J_1$ = 113 MHz, $J_2$ = 59 MHz, $J_3$ = 88 MHz. Signals corresponding to Fe's excited spin state |↓⟩ are small and fall outside the frequency range shown here. ESR conditions in **b,c**: $V_{DC}$ = 50 mV, $I_{DC}$ = 20 pA, $V_{RF}$ = 30 mV.

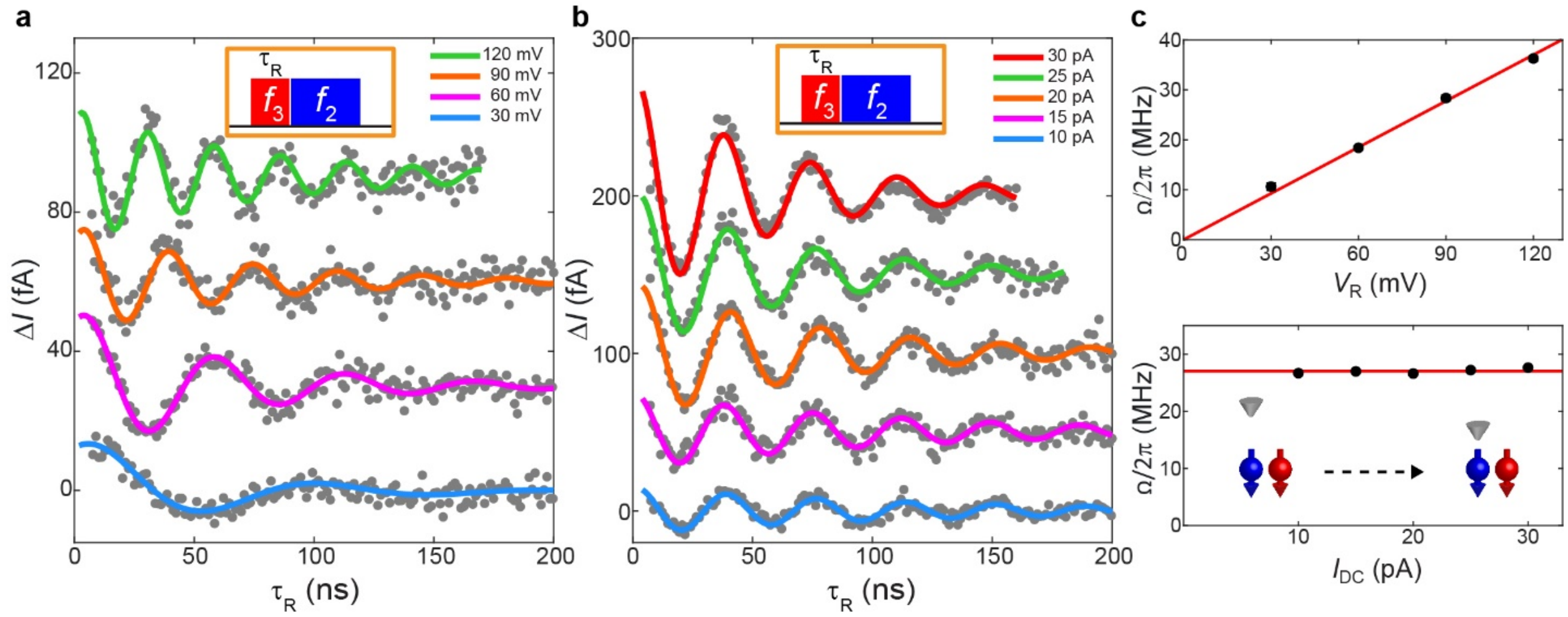

**Extended Data Fig. 3 | Rates of controlled rotations of a remote qubit measured with varying parameters in the two-qubit structure shown in Fig. 1b. a,** Controlled rotations of the remote qubit measured with varying RF driving voltage $V_{R}$ (that acts on the remote qubit). The tunnel current is fixed at $I_{DC}$ = 20 pA. **b,** Controlled rotations of the remote qubit measured at $V_{R}$ = 120 mV with varying tunnel current $I_{DC}$. **c**, Extracted rates of controlled rotations, $\Omega/2\pi$, as functions of the RF driving voltage $V_{R}$ (upper) and tunnel current $I_{DC}$ (lower). Same as Fig. 3b of the main text. All curves in **a**,**b** except the blue ones are vertically shifted for clarity. ESR conditions in **a**: $V_{DC}$ = 50 mV, $I_{DC}$ = 20 pA, $V_{S}$ = 30 mV, $\tau_{S}$ = 200 ns. **b**: $V_{DC}$ = 50 mV, $V_{S}$ = 30 mV, $V_{R}$ = 120 mV, $\tau_{S}$ = 200 ns.

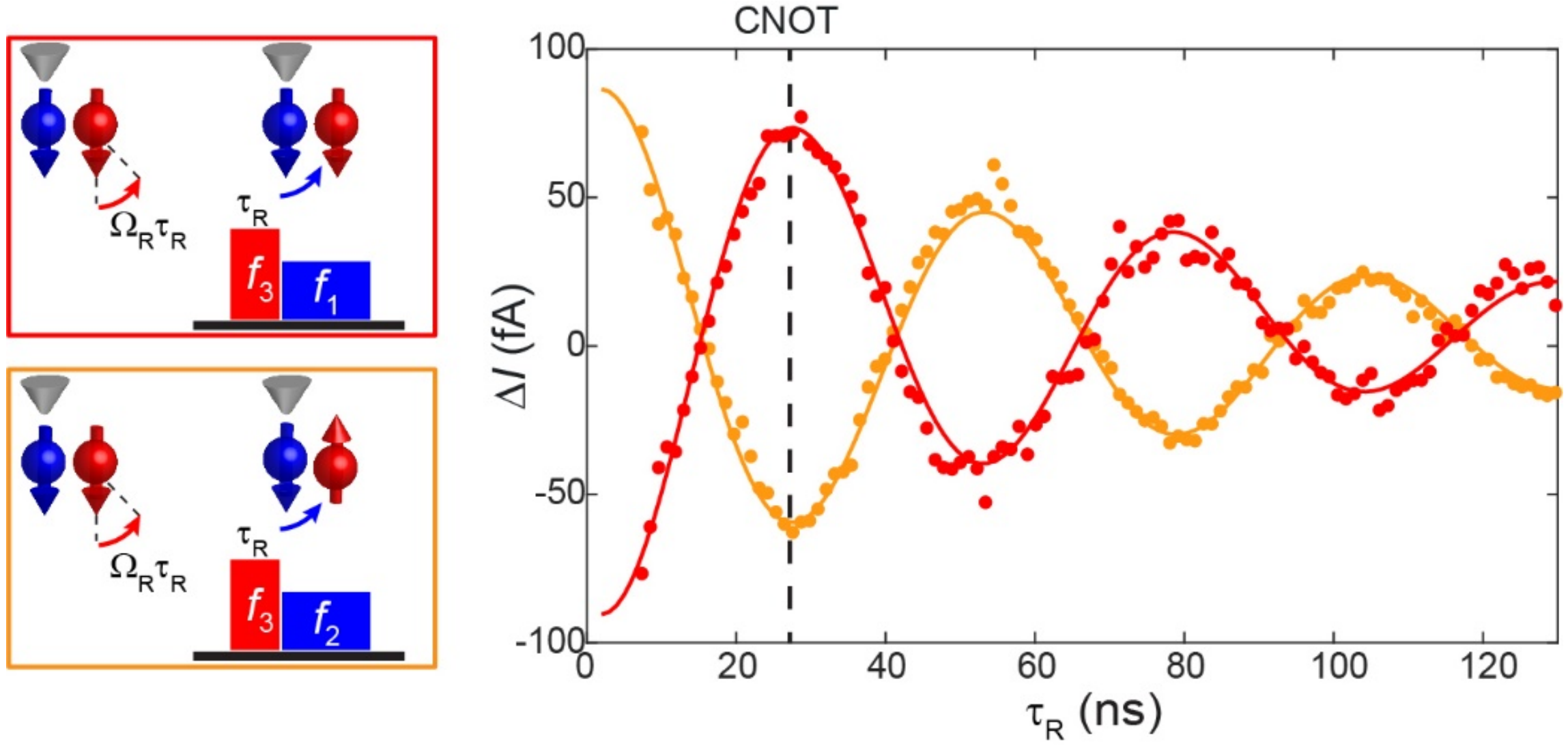

**Extended Data Fig. 4 | Controlled rotations of the remote qubit measured with two different sensing frequencies in the two-qubit structure shown in Fig. 1b.** Left: Schematics of the pulse sequences and corresponding qubit rotations. Right: Controlled rotations of the remote qubit measured with a sensing pulse applied at $f_1$ (red) and $f_2$ (orange). A sensing pulse at $f_1$ excites the transition $|0\rangle|0\rangle \leftrightarrow |1\rangle|0\rangle$, hence effectively detecting state $|0\rangle$ of the remote qubit[3]. In contrast, a sensing pulse at $f_2$ excites the transition $|0\rangle|1\rangle \leftrightarrow |1\rangle|1\rangle$, hence effectively detecting state $|1\rangle$ of the remote qubit. The two oscillations thus show opposite phases. ESR conditions: $V_{DC}$ = 50 mV, $I_{DC}$ = 20 pA, $V_S$ = 30 mV, $V_R$ = 60 mV, $\tau_S$ = 200 ns.

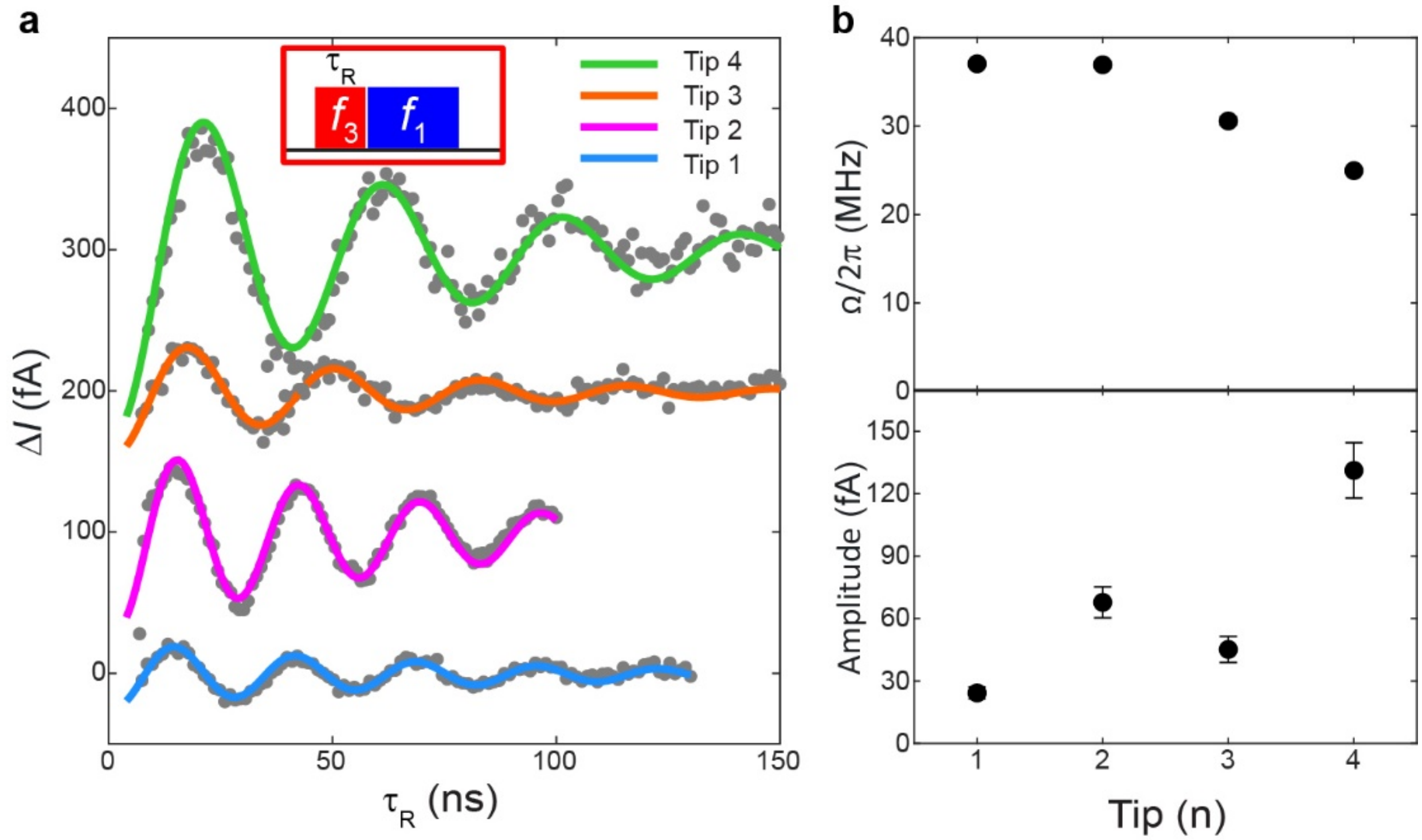

**Extended Data Fig. 5 | Controlled oscillations of the remote qubit measured with different magnetic tips in the two-qubit structure shown in Fig. 1b. a,** Controlled oscillations of a remote qubit measured with different magnetic tips using the same parameters. **b,** Extracted frequencies and amplitudes of the oscillations from data in **a**. Variations of $\Omega/2\pi$ might be related to variations of the tip's effective radius (which affects the radiation range of the RF electric field)[4] or stability of the tip's magnetic moment (see the main text). Green, orange and magenta curves are vertically shifted for clarity. ESR conditions in **a**: $V_{DC}$ = 50 mV, $I_{DC}$ = 20 pA, $V_S$ = 30 mV, $V_R$ = 120 mV, $\tau_S$ = 200 ns.

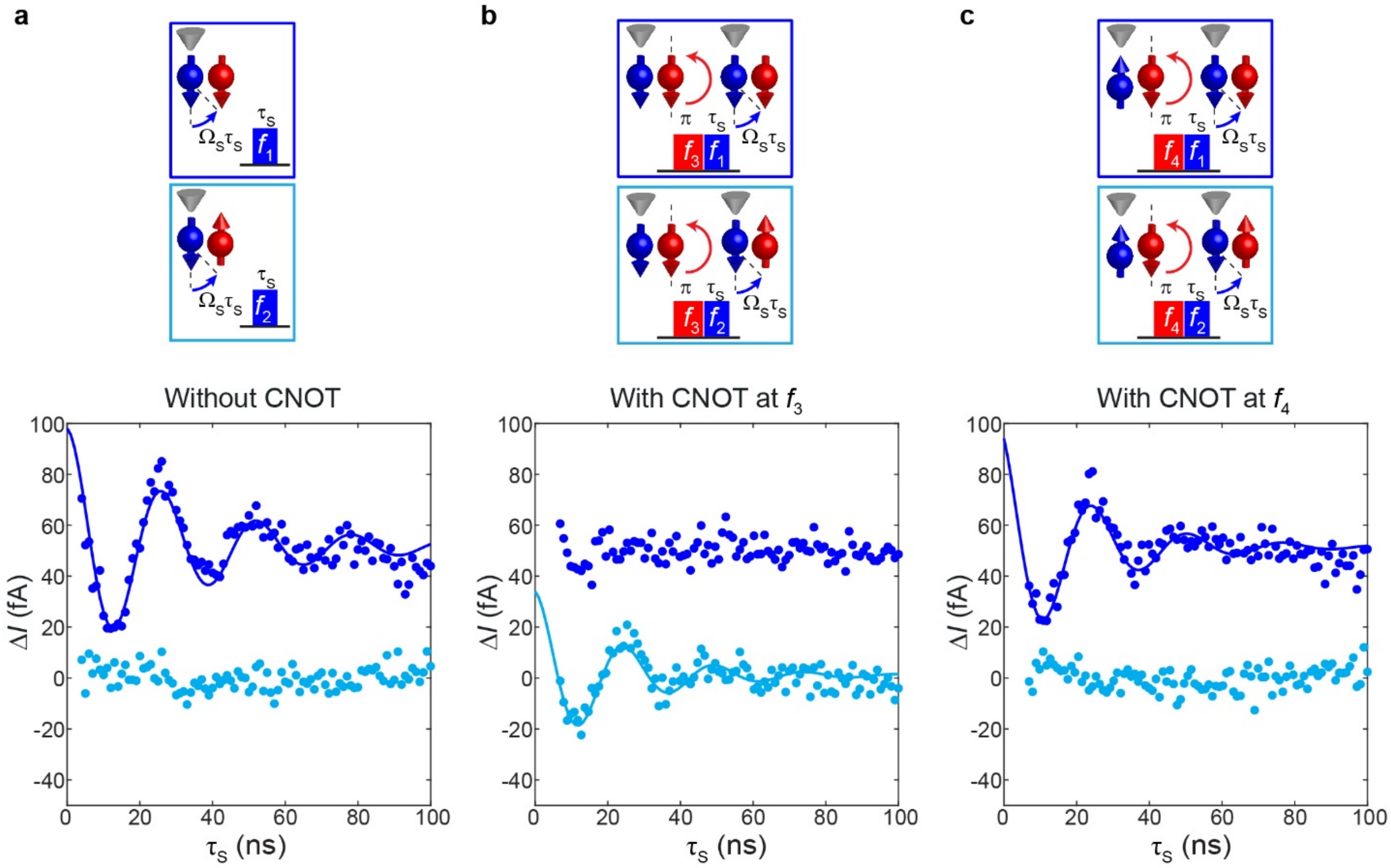

**Extended Data Fig. 6 | Controlled rotation of the sensor qubit without and with applying a CNOT operation of the remote qubit, measured in the two-qubit structure shown in Fig. 1b.** Blue (cyan) curves measured with $f_S = f_1$ ($f_2$) correspond to the controlled rotations of the sensor qubit when the remote qubit is in state $|0\rangle$ ($|1\rangle$). **a**, Controlled rotations of the sensor qubit without any operations on the remote qubit. Same as main Fig. 3c. **b**, Controlled rotations of the sensor qubit with a CNOT operation on the remote qubit at $f_R = f_3$, same as main Fig. 3d. **c**, Similar to **b** but with a CNOT operation on the remote qubit at $f_R = f_4$. In **b**, the CNOT operation on the remote qubit at $f_R = f_3$ transfers the predominant thermal population from state $|0\rangle|0\rangle$ to state $|0\rangle|1\rangle$, thus causing the opposite trends of oscillations in **a** and **b**. In contrast, in **c**, the CNOT operation on the remote qubit at $f_R = f_4$ only affects the minor thermal populations in states $|1\rangle|0\rangle$ and $|1\rangle|1\rangle$, thus showing no obvious changes between **c** and **a**. These results highlight the selective, controlled nature of the CNOT operations. Blue curves are shifted vertically by 50 fA for clarity. ESR conditions: **a**: $V_{DC}$ = 20 mV, $I_{DC}$ = 7.5 pA, $V_S$ = 100 mV. **b,c**: $V_{DC}$ = 20 mV, $I_{DC}$ = 7.5 pA, $V_S = V_R$ = 100 mV.

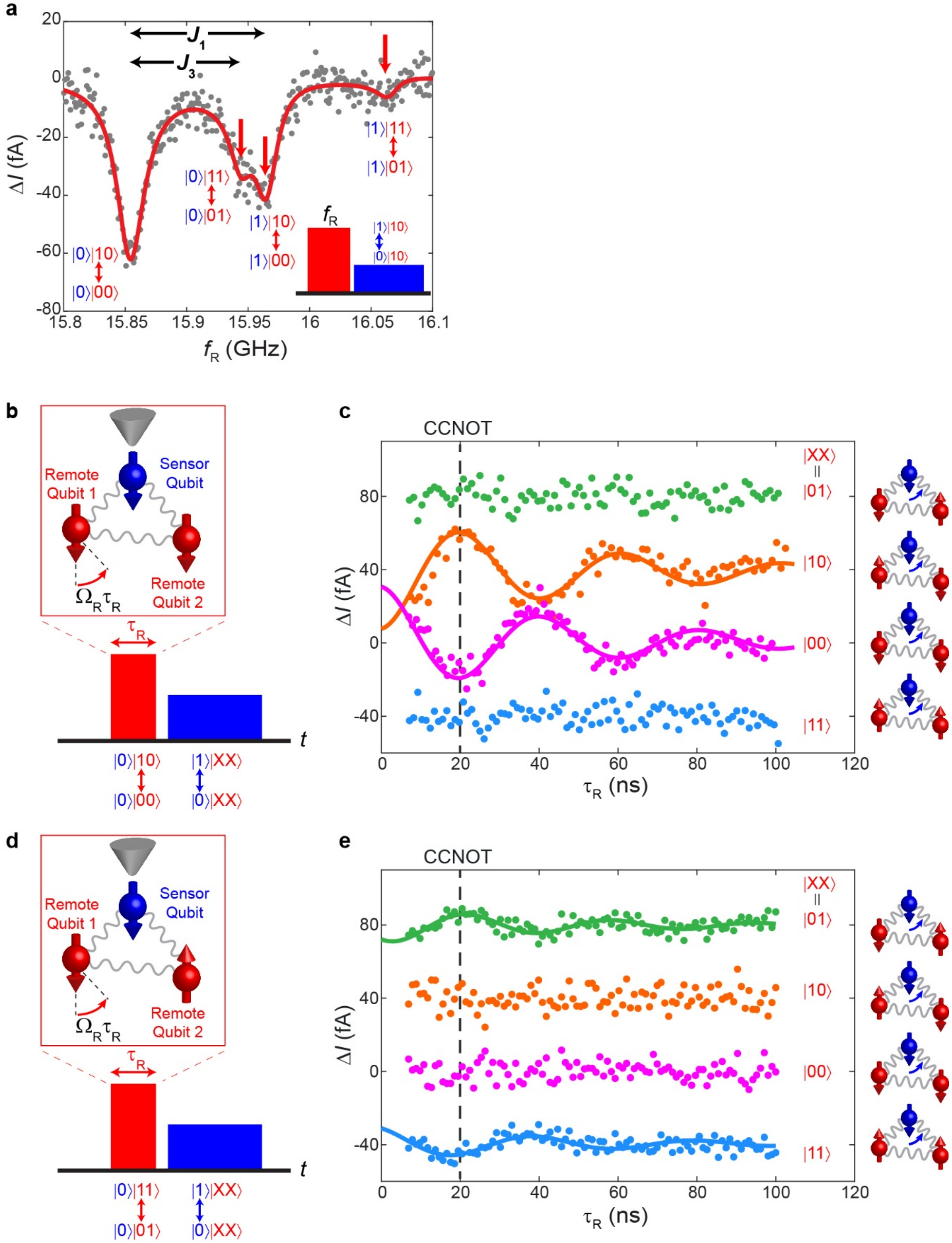

**Extended Data Fig. 7 | Controlled-controlled rotations of a remote qubit in the three-qubit structure shown in Fig. 1c. a,** ESR spectrum of remote qubit 1 measured with the tip placed on the sensor qubit. The frequency $f_R$ of the remote control pulse is swept while the sensing pulse frequency is fixed at the transition $|0\rangle|10\rangle \leftrightarrow |1\rangle|10\rangle$. This spectrum is used for identifying the frequencies of the remote qubit pulse used in this figure and main Fig. 4. **b,** Schematic showing the control scheme of a CCNOT operation, $|0\rangle|00\rangle \leftrightarrow |0\rangle|10\rangle$, that acts on remote qubit 1. Same as main Fig. 4a. **c**, CCNOT operation of remote qubit 1 by driving the transition $|0\rangle|00\rangle \leftrightarrow |0\rangle|10\rangle$ in the control pulse (red pulse in **c**). Four different frequencies of the sensor qubit are used for readout of the four remote qubit states (red kets). Same as Fig. 4b. **d**,**e**, Same as **b**,**c** but with a CCNOT operation at a different transition, $|0\rangle|01\rangle \leftrightarrow |0\rangle|11\rangle$, of remote qubit 1. The CCNOT operation in **b** ($|0\rangle|00\rangle \leftrightarrow |0\rangle|10\rangle$) results in oscillating populations of remote qubits' $|00\rangle$ or $|10\rangle$ states as shown in **c**. The CCNOT operation in **d** ($|0\rangle|01\rangle \leftrightarrow |0\rangle|11\rangle$) results in oscillating populations of remote qubits' $|01\rangle$ or $|11\rangle$ states as shown in **e**. The oscillation amplitude in **e** is smaller due to the smaller thermal populations in remote qubit states $|01\rangle$ and $|11\rangle$. The sinusoidal fits in **c** and **e** show a CCNOT operation time of ~20 ns. Green, orange and blue curves are vertically shifted for clarity. ESR conditions in **a:** $V_{DC}$ = 50 mV, $I_{DC}$ = 20 pA, $V_S = V_R$ = 30 mV, $\tau_S = \tau_R$ = 200 ns. **c,e**: $V_{DC}$ = 50 mV, $I_{DC}$ = 20 pA, $V_S$ = 50 mV, $V_R$ = 80 mV, $\tau_S$ = 200 ns.

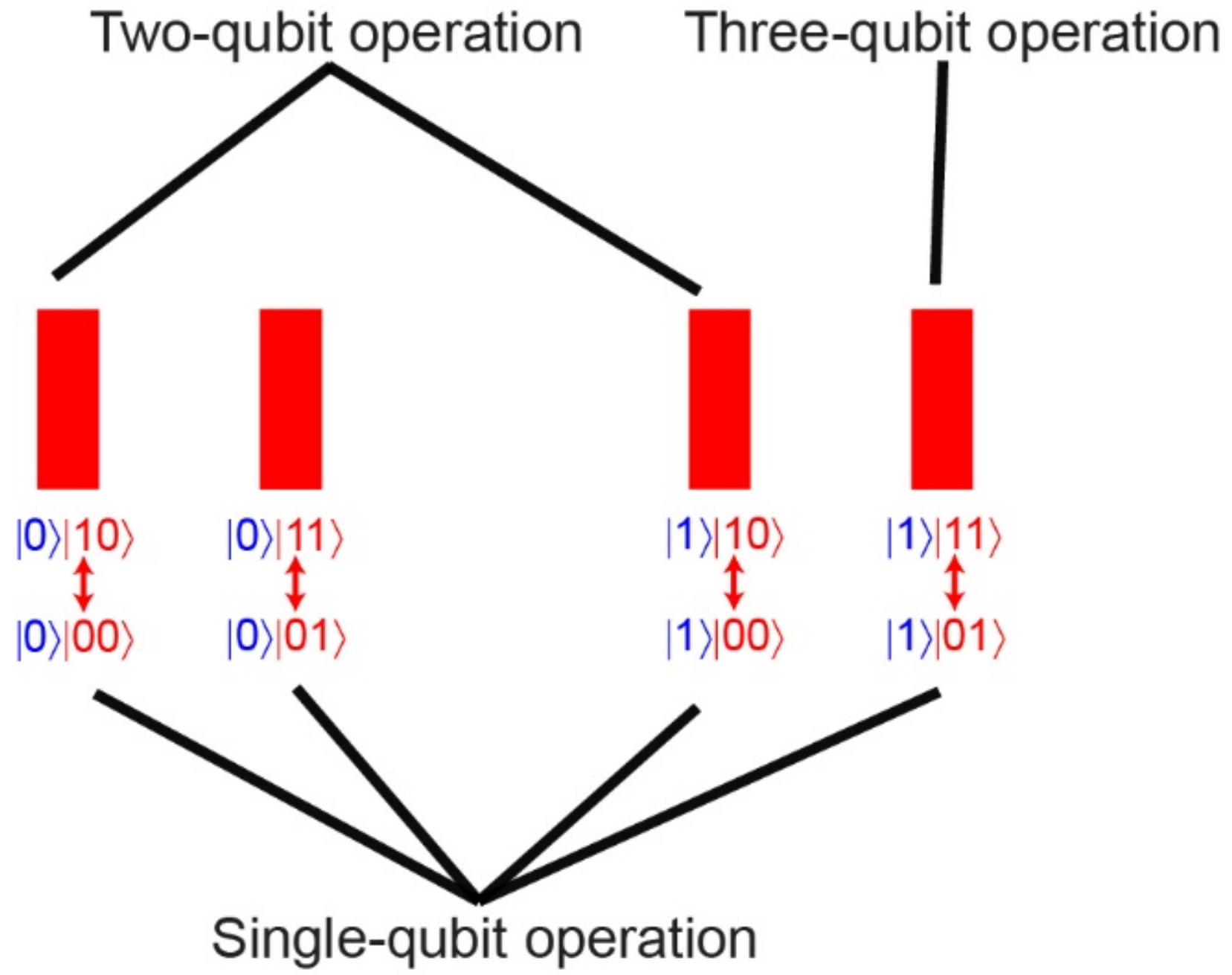

**Extended Data Fig. 8 | Strategy to perform single-, two-, and multi-qubit operations in our platform.** Here for simplicity we use a three-qubit structure as an example. Single-qubit operations are performed by non-selectively exciting all ESR transitions of a qubit with the same strength. Two-qubit operations are performed by selectively exciting ESR transitions that correspond to a specific quantum state of one control qubit. Multi-qubit operations are performed by selectively exciting ESR transitions that correspond to a specific quantum state of multiple control qubits.

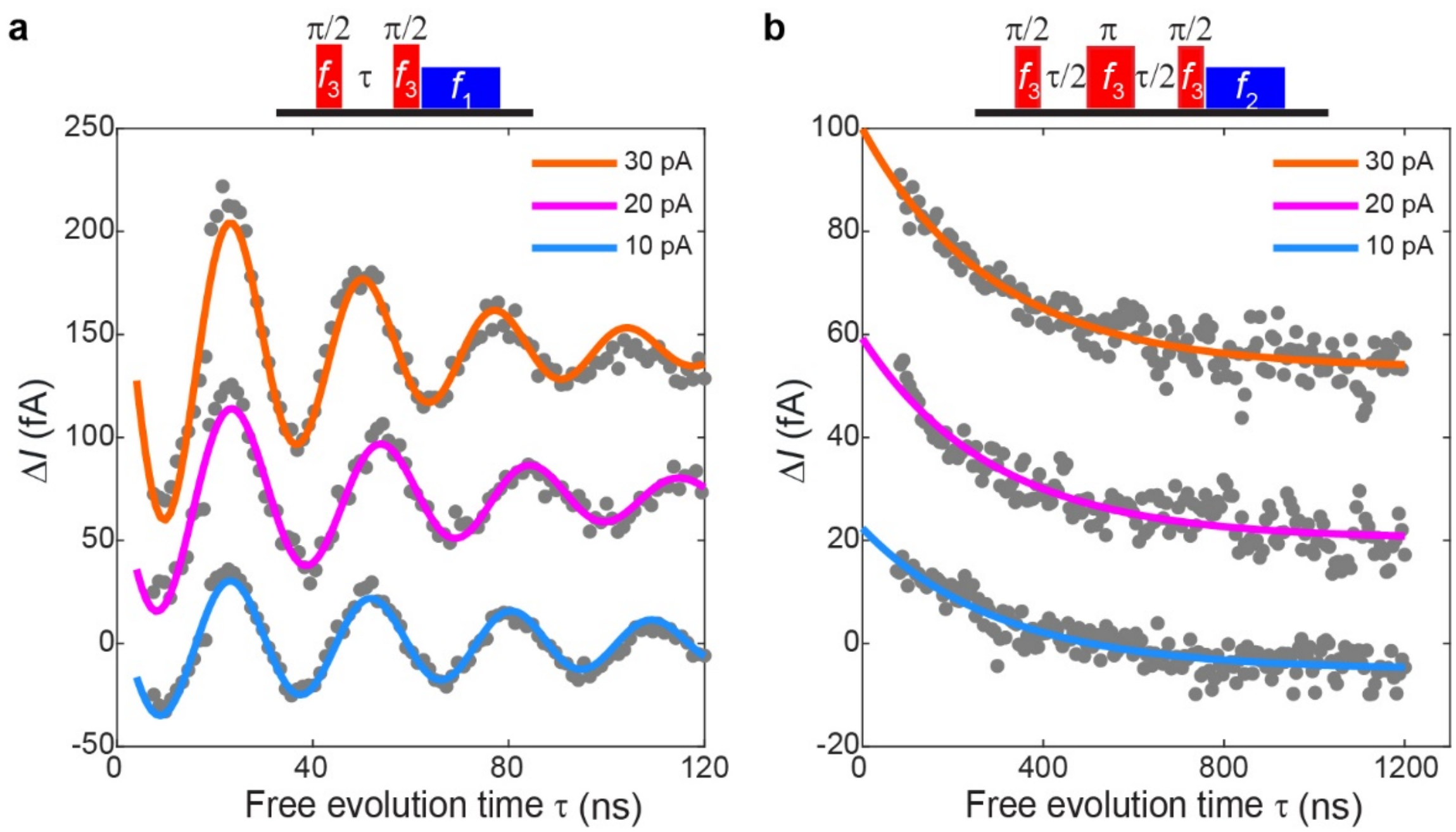

**Extended Data Fig. 9 | Spectra of Ramsey and spin echo measurements corresponding to the data points shown in the insets of Fig. 5b and 5c.** Orange and magenta curves are vertically shifted for clarity. The blue curve in **a** is the same as the spectrum shown in main Fig. 5b. The blue curve in **b** is the same as the spectrum shown in main Fig. 5c.